\documentclass[preprint,aps]{revtex4}

\usepackage{epsfig}
\usepackage{graphicx}
\usepackage{color}
\usepackage{amssymb}

\begin{document}

\newcommand{\be}{\begin{equation}}
\newcommand{\ee}{\end{equation}}
\newcommand{\bea}{\begin{eqnarray}}
\newcommand{\eea}{\end{eqnarray}}
\newcommand{\nn}{\nonumber}
\newcommand{\de}{\partial}

\def\a{\alpha}
\def\b{\beta}
\def\d{\delta}        \def\D{\Delta}
\def\e{\epsilon}
\def\eps{\varepsilon}
\def\f{\phi}          \def\F{\Phi}
\def\vf{\varphi}
\def\g{\gamma}        \def\G{\Gamma}
\def\h{\eta}
\def\i{\iota}
\def\j{\psi}          \def\J{\Psi}
\def\k{\kappa}
\def\lam{\lambda}   \def\L{\Lambda}
\def\m{\mu}
\def\n{\nu}
\def\o{\omega}   \def\O{\Omega}
\def\p{\pi}      \def\P{\Pi}
\def\q{\theta}   \def\Q{\Theta}
\def\s{\sigma}   \def\S{\Sigma}
\def\t{\tau}
\def\u{\upsilon}  \def\U{\Upsilon}
\def\x{\xi}      \def\X{\Xi}
\def\z{\zeta}

\linespread{1.5}

\title{Electron-impact resonant vibrational excitation and dissociation processes involving vibrationally excited N$_2$ molecules}

\author{
V. Laporta$^{1,2,*}$\footnote[0]{$^*$ vincenzo.laporta@ba.imip.cnr.it},
D.A. Little$^2$,
R. Celiberto$^{3,1}$
and J. Tennyson$^{2}$
}

\affiliation{$^1$ Istituto di Metodologie Inorganiche e dei Plasmi, CNR, Bari, Italy}
\affiliation{$^2$ Department of Physics and Astronomy, University College London, London WC1E 6BT UK}
\affiliation{$^3$ Dipartimento di Ingegneria Civile, Ambientale, del Territorio, Edile e di Chimica, Politecnico di Bari, Italy}

\begin{abstract}
  Resonant vibrational excitation cross sections and the corresponding
  rate coefficients for electron-N$_2$ collisions occurring through
  the N$_2^-(\textrm{X}\ ^2\Pi_g)$ resonant state are reviewed. New
  calculations are performed using accurate potential energies curves
  for the N$_2$ electronic ground state, taken from literature, and
  for the N$_2^-$ resonant state, obtained from $R$-matrix
  calculations. The calculations are extended  to  resonant
  excitation processes involving the N$_2$ ground state vibrational
  continuum, leading to dissociation. Electron impact dissociation is
  found to be significant from higher vibrational levels. Accurate
  analytical fits for the complete set of the rate coefficients are
  provided. The behavior of the dissociative cross sections is
  investigated for rotationally excited N$_2$ molecules, with
  $J=50,100$ and 150 and for different vibrational levels.
\end{abstract}

\maketitle

\section{Introduction \label{sec:intro}}
Nitrogen molecule plays a role of fundamental importance in many scientific and industrial activities. Typical examples are provided by air plasmas studied in a variety of fields such as environmental research, Earth's atmosphere phenomena, combustion, and aerospace technologies~\cite{Gordillo_Vazquez_2008, Shang_Surzhikov,   Laporta_Bruno_2013, Capitelli_et_al_ICPIG_XXXI}. Detailed chemistries of processes involving molecular nitrogen have been prepared for such studies~\cite{PSST.22.025008,ApJ2013}.

One of the main aspects in the formulation of a model for non-equilibrium, nitrogen-containing plasmas is represented by the description of the vibrational kinetic and its role in redistributing the internal energy of the plasma among the atomic and molecular degrees of freedom. Electron-molecule collisions, involving vibrationally excited N$_2$ molecules and leading to vibrational excitations, represent a central process in the kinetic evolution of the plasma. In particular, resonant vibrational excitation (RVE) process, which occurs \emph{via} the capture of the incident electron by the molecule with the formation of an unstable molecular anion, is one of the most important processes. In fact, decay of this resonance state can lead efficiently to single and multi-quantum vibrational excitations, which can strongly affect the vibrational population of the species in the plasma.

Several experimental~\cite{0953-4075-29-6-023, Ristic2007410, Allan_1985_N2} and theoretical~\cite{PhysRevLett.43.1926, PhysRevA.36.1632, 0022-3727-32-20-307} works have been devoted to study vibrational excitation of nitrogen by electron-impact. A complete set of cross sections and related rate coefficients, were recently  reported in Ref.~\cite{0963-0252-21-5-055018} (hereafter referred to as I), for the RVE process: \be e(\epsilon) + \textrm{N}_2(\textrm{X}\ ^1\S^+_g;v)\to \textrm{N}_2^-(\textrm{X}\,^2\P_g) \to e(\epsilon') + \textrm{N}_2(\textrm{X}\ ^1\S^+_g;v')\,, \label{eq:RVE_process}\ee where the incident electron with energy $\epsilon$ is captured by the N$_2$ molecule, initially in its ground electronic state $\textrm{X}\,^1\S^+_g$ and in the vibrational level $v$, with the formation of the resonant state $\textrm{N}_2^-(\textrm{X}\ ^2\P_g)$ which decays into a free electron, with energy $\epsilon'$, and a vibrationally excited molecule $\textrm{N}_2(\textrm{X}\,^1\S^+_g;v')$. Cross sections $\sigma_{v,v'}(\epsilon)$ for process (\ref{eq:RVE_process}) were calculated in I using the so-called \textit{local-complex-potential} (LCP) model for the scattering description, using Morse-like potential energies curves~\cite{Gilmore1965369} as input parameters, for both N$_2(\textrm{X}\ ^1\S^+_g)$ and N$_2^-(\textrm{X}\,^2\P_g)$ states. The width for the N$_2^-(\textrm{X}\,^2\P_g)$ state was based on a semi-empirical analytical function optimized to reproduce the experimental data.

In this paper we provide new RVE cross section calculations for process (\ref{eq:RVE_process}) performed using an accurate potential curve for the N$_2$ molecule~\cite{:/content/aip/journal/jcp/125/16/10.1063/1.2354502}. For the N$_2^-$ ion, the potential curve is obtained by \textit{ab initio} calculations using the \textit{R}-matrix method which also provides the resonance width as a function of the internuclear distance. The cross section calculations are extended to the RVE process ending in the vibrational continuum of the ground state~\cite{0953-4075-26-21-007}. The repulsive nature of the curve induces the separation of the nuclei so that the molecule undergoes dissociation with the production of two stable nitrogen atoms in their lowest electronic state. The resonant dissociation process can then be represented as: \be e(\epsilon) + \textrm{N}_2(\textrm{X}\,^1\S^+_g;v,J)\to \textrm{N}_2^-(\textrm{X}\,^2\P_g)\to e(\epsilon') + \textrm{N}_2(\textrm{X}\,^1\S^+_g;continuum) \to e(\epsilon') + 2\, \textrm{N}(^4\textrm{S})\,.\label{eq:Rdiss_process}\ee This dissociative channel is particularly important in plasma kinetics because, in non-equilibrium conditions, it can   compete with the dissociation from heavy species  collisions~\cite{:/content/aip/journal/jcp/138/4/10.1063/1.4774412,  Esposito20061, :/content/aip/journal/pof2/26/4/10.1063/1.4872336}.

The cross sections for processes (\ref{eq:RVE_process}) and (\ref{eq:Rdiss_process}) have been obtained as a function of the incident electron energy and for all the initial vibrational levels $v$ by adopting the LCP model, and then used for the calculation of the corresponding rate coefficients, assuming a Maxwell distribution for electrons, according to the equation:
\begin{equation}
\kappa_{v,v'}(T) = \sqrt{\frac{8}{m_e\,\pi}}\left(\frac{1}{T}\right)^{3/2}\int^\infty _{\epsilon_{th}} d\epsilon\ \epsilon\cdot e^{-\frac{\epsilon}{T}}\cdot \sigma_{v,v'}(\epsilon)\,, \label{eq:rate coefficient}
\end{equation}
where $m_e$ is the electron mass and the temperature $T$ is expressed in energy units. It worth be noted that Eq.~(\ref{eq:rate coefficient}) is no longer valid in non-equilibrium condition. Cross sections are also  studied for some values of the rotational quantum number, namely $J=50,$ 100 and 150.

The paper is organized as follows: in the next section the \textit{R}-matrix calculations are described and the main equations of the LCP model are shown. The results are presented and commented in section~\ref{sec:results} while, in section~\ref{sec:conc}, a brief summary concludes the work.

\section{Theoretical model \label{sec:PEC}}

\subsection{\textit{R}-matrix method\label{sec:Rmatrix}}
Electron-N$_2$ calculations were performed using the \textit{R}-matrix method as implemented in the UKRMol codes~\cite{jt518}. For details of this methodology we refer to the review by one of us~\cite{Tennyson201029}. Put simply, the \textit{R}-matrix method divides space into an inner region defined by a sphere centered on the target center-of-mass. This sphere, here taken to be 10~$a_0$, is assumed to enclose the entire charge cloud of the $N$-electron target. Within the sphere the wave function of the ($N$+1)-electron scattering problem is built from target wave functions and extra functions designed to represent the scattering continuum. Here, and in general, this problem is built about using complete active space (CAS) configuration interaction (CI) representation of the target wave function for which a particularly efficient purpose-built algorithm is used~\cite{jt180}. In the outer region, the interaction of the scattering electron with the target is assumed to occur only \emph{via} diagonal and off-diagonal multipole moments of the target. While the inner region problem only has to be solved once for each total scattering symmetry, the much faster outer region problem is solved at each scattering energy of interest. Below we give specific details for the present calculation.

Target calculations used the cc-pVQZ Gaussian Type Orbital (GTO) basis
set due to Dunning. Orbitals for the N$_2$ target where generated
using multi-configuration self-consistent field (MCSCF) calculations
run in MOLPRO~\cite{MOLPRO_brief}. The CAS used in these calculations
and to define the target wave function in the \textit{R}-matrix
calculations is given by: \[{(\rm 1 \sigma_g, 1 \sigma_u)}^4{( 2
  \sigma_g, 2 \sigma_u, 1\pi_u, 3\sigma_g,1\pi_g,3\sigma_u
  )}^{10}\,.\] A total of 128 target states were generated (eight per
symmetry) of which the lowest 49 in energy were retained for the inner
region calculation. Calculations were performed for a 100 geometries
from 0.8~\AA\ to 3.77~\AA\ in steps of 0.03~\AA.

For the scattering calculations, $(4 \sigma_g, 5 \sigma_g, 4 \sigma_u, 2\pi_u, 2\pi_g, 1\delta_g)$ target orbitals were retained. These were augmented by continuum orbital containing up to g ($l=4$) functions represented by a GTO expansion at the target center-of-mass~\cite{jt286}. These were orthogonalized to the target orbitals with a deletion threshold set to $10^{-7}$~\cite{Morgan1998120}. The target times continuum configurations were augmented by the following short-range functions based on the use of target orbitals: \[ {(\rm 1 \sigma_g, 1 \sigma_u)}^4{( 2 \sigma_g, 2 \sigma_u, 1\pi_u,  3\sigma_g,1\pi_g,3\sigma_u )}^{11}\,, \] which involves placing the scattering electron in the target CAS, and \[ {(\rm 1 \sigma_g, 1   \sigma_u)}^4{( 2 \sigma_g, 2 \sigma_u, 1\pi_u,  3\sigma_g,1\pi_g,3\sigma_u)}^{10}{(2\pi_u, 4 \sigma_g, 5 \sigma_g,   2\pi_g, 1\delta_g, 4 \sigma_u)}^{1}\,, \] where the scattering electron enters and otherwise unoccupied target virtual. These configurations were not constrained by contracting to target CI wave functions~\cite{jt189}.

For computational efficiency, only the lowest two target states, $\rm X~^1\Sigma^{+}_{g}$ and $\rm A~^3\Sigma^{+}_{u}$, were retained in the outer region calculation. This problem was solved by propagating the \textit{R}-matrix to 100.1~$a_0$ and then using a Gailitis expansion. The resulting eigenphases were searched for a resonance which was fitted to a Breit-Wigner form using an automated procedure~\cite{Tennyson1984421}. These calculations concentrated on the $^2\Pi_g$ total symmetry as this is the symmetry of the well known, low-lying N$_2^-$ shape resonance.

\begin{center}
\begin{figure}
\includegraphics[scale=.7]{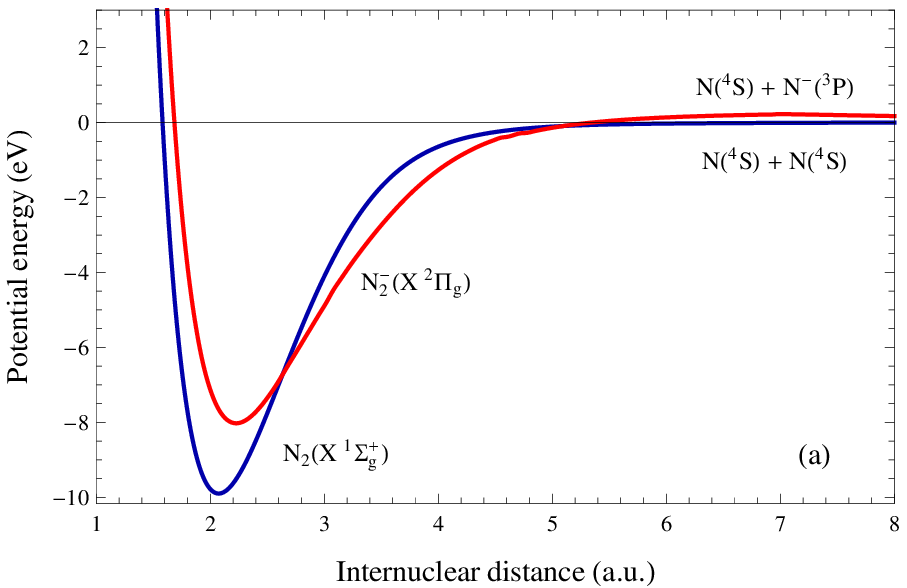}\hspace{1cm}
\includegraphics[scale=.7]{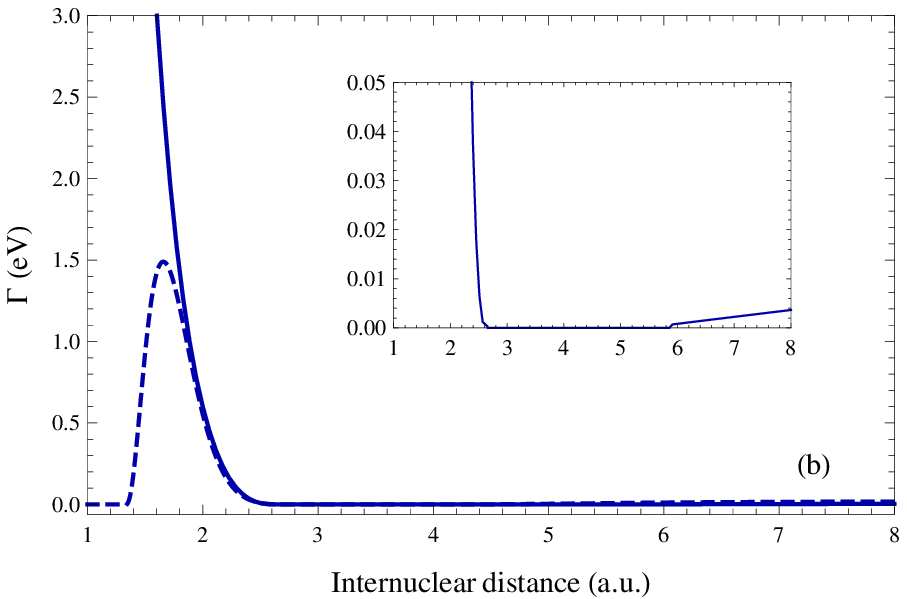}
\caption{(a) Potential energy curves for N$_2(\mathrm{X}\ ^1\Sigma_g^+)$ and N$_2^-(\mathrm{X}\ ^2\Pi_g)$ and
(b) resonance width $\Gamma(R)$ as a function of the internuclear distance: solid line, this work;
dashed line, phenomenological width calculated in I.  The inset plot shows a magnified view of our
resonance width ($\Gamma(R)\leq 0.05$ eV). \label{fig:pes}}
\end{figure}
\end{center}

The N$_2(\textrm{X}\ ^1\Sigma_g^+)$ potential energy curve is taken from Le Roy {\it et al.}~\cite{:/content/aip/journal/jcp/125/16/10.1063/1.2354502} who obtained it by an accurate fit to spectroscopic data. The resulting potential curves for both N$_2$ and N$_2^-$, along with the corresponding width $\Gamma(R)$, are shown in Fig.~\ref{fig:pes}(a)-(b) respectively. The $\rm N^-_2({X}~^{2}{\Pi}_{g})$ resonance curve crosses $\rm N_2(X~^1\Sigma^{+}_{g})$ at $\sim2.657$ $a_0$ and $\sim5.132$ $a_0$. In this intermediate region the molecular ion N$_2^-$ becomes stable and the resonance width vanishes, as is shown in the inset box in Fig.~\ref{fig:pes}(b). For geometries where N$_2^-$ is bound, the position of the bound state was determined using the same model and by performing negative energy scattering calculations~\cite{jt106}. Table~\ref{tab:spectr_data} reports some relevant spectroscopical parameters for the two potentials.

\begin{center}
\begin{table}
\begin{tabular}{ccc}
\hline
               & $\textrm{N}_2(\textrm{X}\ ^1\S^+_g)$ & $\textrm{N}^-_2(\textrm{X}\ ^2\P_g)$\\
  \hline
  $R_e$ (a$_0$) & 2.07 & 2.23 \\
  $D_e$ (eV)   & 9.89 & 8.22 \\
  $T_e$ (eV)   &  0   & 1.97 \\
  RP    (eV)   &  --  & 2.34 \\
  EA (eV)      &  --  & -- 0.30 \\
  \hline
\end{tabular}
\caption{Equilibrium distance ($R_e$), dissociation energy ($D_e$), vertical excitation ($T_e$), resonance position calculated at the N$_2$ equilibrium bond-length (RP) and electronic affinity (EA) for the N$_2$ and N$_2^-$ ground state potential energy curves. \label{tab:spectr_data}}
\end{table}
\end{center}

\subsection{Vibrational dynamics \label{sec:Thmodel}}
In this section the main equations for the description of the vibrational dynamics of the collision in the framework of the LCP model are summarized. Extensive theoretical formulations of electron-molecule resonant scattering can be found elsewhere~\cite{0034-4885-31-2-302, Domcke199197, wadehra}.

The cross sections for the RVE processes in~(\ref{eq:RVE_process}),
with incoming electron energy $\e$, were calculated using the
following formula: \be \s_{v,v'}(\e) = g\,
\frac{64\,\pi^5\,m^2}{\hbar^4}\, \frac{k'}{k}\, |\mathcal
T_{v,v'}(\e)|^2\,, \label{eq:xsec}\ee where $k$($k'$) is the ingoing
(outgoing) electron momentum, $g$ contains the spin-statistic weight
factors and $v$($v'$) represents the bound initial (final) nitrogen
vibrational levels. $\mathcal T_{v,v'}$ is the \textit{T}-matrix of
the process. The definition in Eq.~(\ref{eq:xsec}) can be extended to
the resonant dissociation process (\ref{eq:Rdiss_process}), by
considering that the final vibrational energy falls now in the
continuum spectrum of the N$_2$ ground state potential. The right-hand
side of Eq.~(\ref{eq:xsec}) retains  the same form but it
expresses now the energy-differential cross section
${d\s_{v,\epsilon'}(\e)}/{d\e'}$. An extra integration over the final
continuum levels of energy $\e'$ is thus required according
to~\cite{0953-4075-26-21-007, SandT_NJP, Celiberto_et_al_PRA_1999}:
\be \s_v(\e) =
\int_{\e_{th}}^{\e_{max}}\,d\e'\,\frac{d\s_{v,\epsilon'}(\e)}{d\e'}\,,
\label{eq:cont_xsec}\ee where $\e_{th}$ is the dissociation threshold.
In our calculations the integration over the continuum has been
extended up to $\e_{max} = \e_{th}+10$~eV.

According to the LCP model of Fano-Bardsley's theory of resonant scattering~\cite{Bardsley_1968, 0034-4885-31-2-302} the $\mathcal T$-matrix in Eq. (\ref{eq:xsec}) is given by: \be \mathcal T_{v,v'}(\e) = \langle \chi_{v'}|V_{dk'}|\xi \rangle\,, \ee where $\xi(R)$ is the resonant state nuclear wave function, solution of the Schrodinger-like equation \be (T_N+V^--\frac i 2\G-E)\xi(R) = -V_{dk}\,\chi_v\,, \ee with total energy $E=\e+\e_v$ and for the resonant complex potential $(V^-(R),\G(R))$. Moreover, $\chi_{v(v')}(R)$ is the initial (final) vibrational wave function, with the corresponding eigenvalues $\e_{v(v')}$, belonging to the N$_2$ ground state potential energy, denoted by $V^0(R)$, and $T_N$ is the nuclear operator. Finally, $V_{dk}$ is the continuum-discrete coupling potential given by, \be V_{dk} = \sqrt{\frac 1{4\pi} \frac{\G(R)}{2\pi} \frac{\hbar^2}{m\,k(R)}}\,, \ee where $k(R) = \sqrt{2m\left[V^-(R)-V^0(R)\right]}/\hbar$.

\section{Results and discussion \label{sec:results}}
\begin{figure}
\includegraphics[scale=.7]{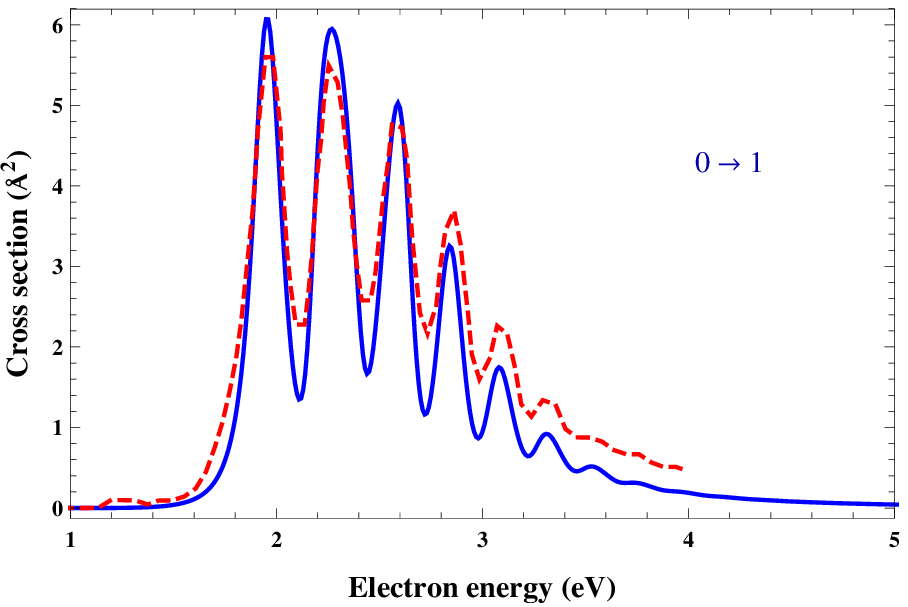}\hspace{1cm}
\includegraphics[scale=.7]{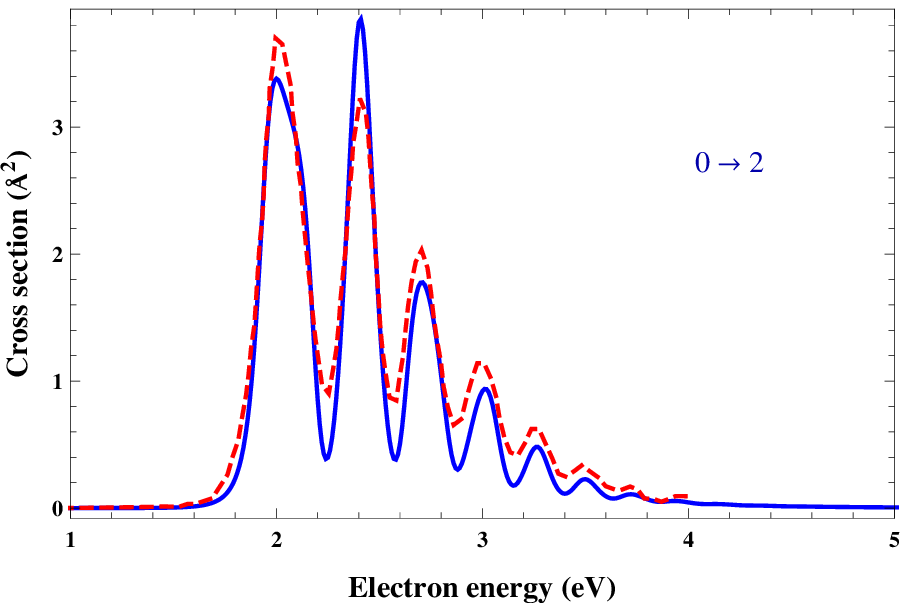}
\\
\includegraphics[scale=.7]{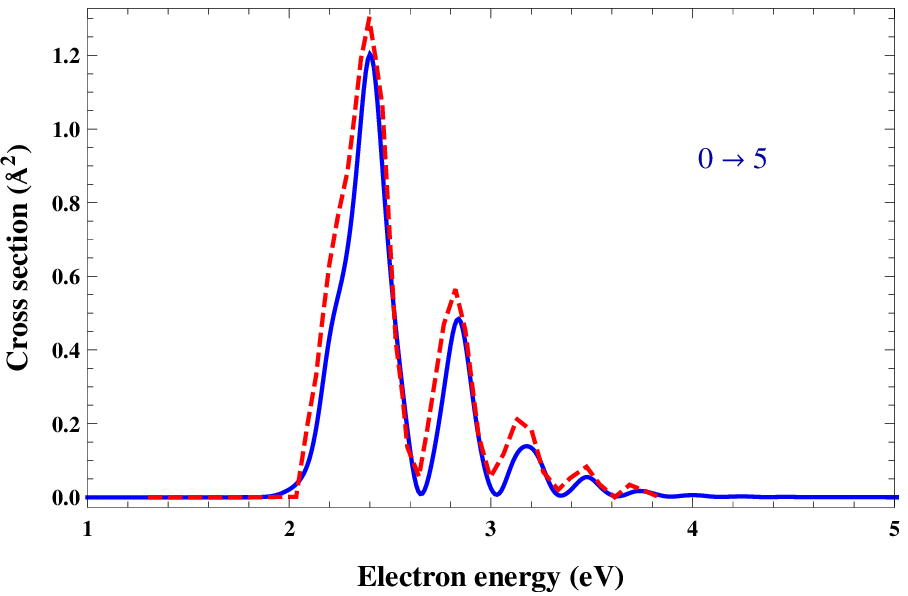}\hspace{1cm}
\includegraphics[scale=.7]{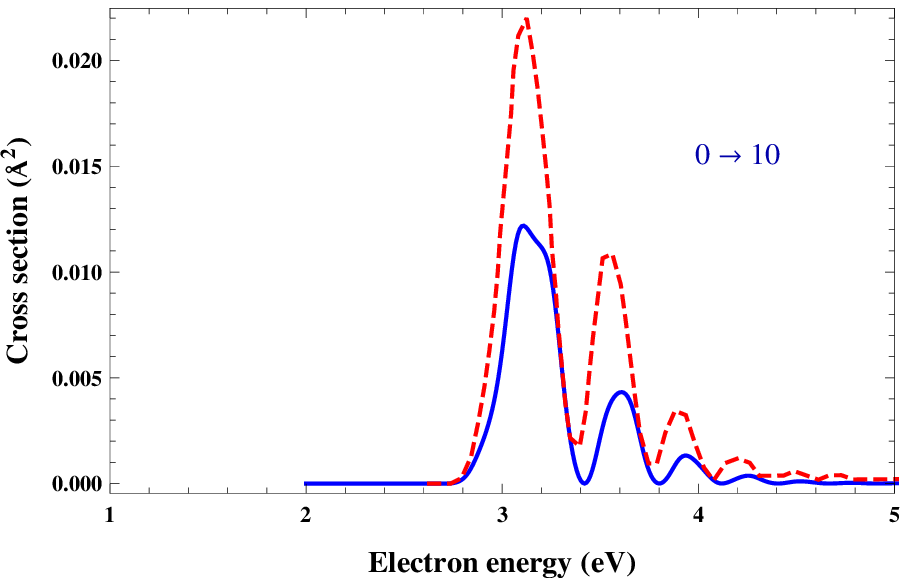}
\caption{Cross section comparison between the present calculations (full-blue line) and the measurements (dashed-red line) of Allan~\cite{Allan_1985_N2}. The experimental data have an estimated uncertainty of $\pm20\%$. \label{fig:Inelastic_comp_Allan}}
\end{figure}

\begin{figure}
\includegraphics[scale=.7]{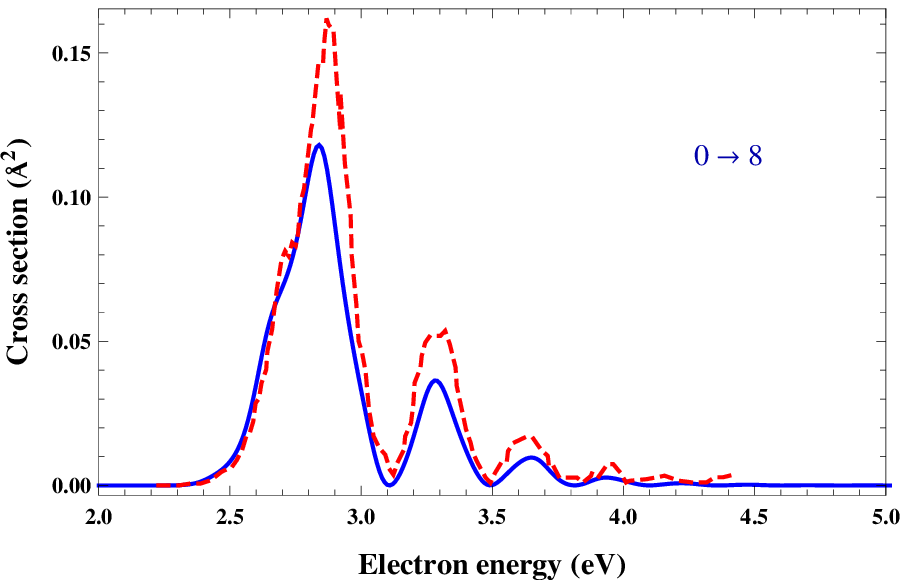}\hspace{1cm}
\includegraphics[scale=.7]{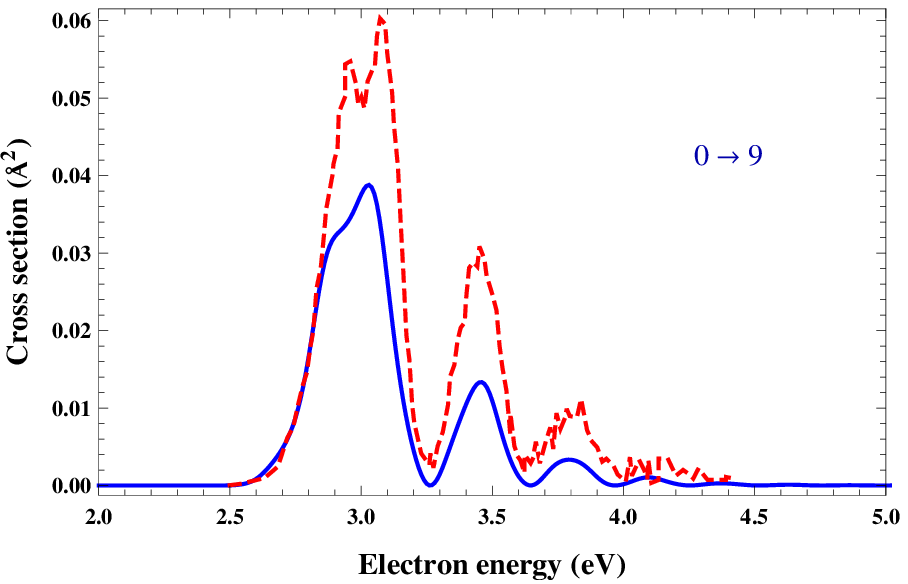}
\caption{Cross section comparison between the present calculations (full-blue line) and the measurements of Vi\'{c}i\'{c} \emph{et al.}~\cite{0953-4075-29-6-023}. \label{fig:comp_vicic}}
\end{figure}

The potential curve for the N$_2$ ground state, used in the present calculations, supports 59 vibrational levels, which are reported in Table~\ref{tab:vib_lev_N2}. This is fewer than the 68 levels found for the Morse curve adopted in I, which implies that there is not a one-to-one correspondence in the energy eigenvalues for the same $v$ in the two sets of calculations. However, the energies of the first few levels do not show large differences compared with those of I and so the updated cross sections remain practically the same. This means that the good agreement between the calculated cross sections with the experimental measurements observed in I is retained in the present results, as can be seen in Figs.~\ref{fig:Inelastic_comp_Allan} and \ref{fig:comp_vicic} where the new theoretical results are compared with the experimental data of Allan~\cite{Allan_1985_N2} and Vi\'{c}i\'{c} \emph{et al.}~\cite{0953-4075-29-6-023} respectively.
\begin{center}
\begin{table}
\begin{tabular}{ccccccccccc}
\hline
 $v$ & $\e_v$ (eV) &~~~~& $v$ & $\e_v$ (eV)    &~~~~& $v$ & $\e_v$ (eV)      &~~~~& $v$ & $\e_v$ (eV) \\
 \hline
 0 &  0.000 && 15 &  3.959 &&  30 &   7.084 &&  45 &   9.163  \\
 1 &  0.288 && 16 &  4.195 &&  31 &   7.260 &&  46 &   9.252  \\
 2 &  0.573 && 17 &  4.426 &&  32 &   7.430 &&  47 &   9.335  \\
 3 &  0.855 && 18 &  4.654 &&  33 &   7.596 &&  48 &   9.409  \\
 4 &  1.133 && 19 &  4.878 &&  34 &   7.757 &&  49 &   9.476  \\
 5 &  1.408 && 20 &  5.099 &&  35 &   7.913 &&  50 &   9.535  \\
 6 &  1.679 && 21 &  5.315 &&  36 &   8.064 &&  51 &   9.587  \\
 7 &  1.947 && 22 &  5.528 &&  37 &   8.210 &&  52 &   9.631  \\
 8 &  2.211 && 23 &  5.737 &&  38 &   8.350 &&  53 &   9.667  \\
 9 &  2.471 && 24 &  5.942 &&  39 &   8.485 &&  54 &   9.696  \\
10 &  2.728 && 25 &  6.143 &&  40 &   8.614 &&  55 &   9.717  \\
11 &  2.982 && 26 &  6.339 &&  41 &   8.737 &&  56 &   9.732  \\
12 &  3.232 && 27 &  6.532 &&  42 &   8.853 &&  57 &   9.742  \\
13 &  3.478 && 28 &  6.721 &&  43 &   8.963 &&  58 &   9.748  \\
14 &  3.720 && 29 &  6.905 &&  44 &   9.067 &&                \\
\hline
\end{tabular}
\caption{Vibrational levels given by the N$_2$  potential energy curve~\cite{:/content/aip/journal/jcp/125/16/10.1063/1.2354502} counted from the lowest level $v=0$, which has a zero point energy of 0.146 eV.\label{tab:vib_lev_N2}}
\end{table}
\end{center}

\begin{center}
\begin{figure}
\includegraphics[scale=.25, angle = -90]{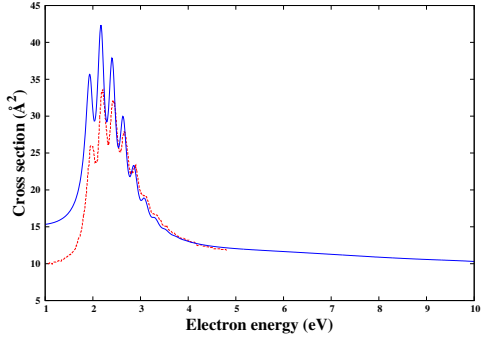}
\caption{Comparison between calculated and measured total cross sections (see text).
Present theoretical results (full-blue line) and experimental data (dashed-red line)~\cite{Kennerly_1980,Itikawa_2006}. \label{fig:Elastic_comparison}}
\end{figure}
\end{center}

Figure~\ref{fig:Elastic_comparison} compares also the calculated total  cross sections with the experimental data of Refs.~\cite{Kennerly_1980,Itikawa_2006}. Our theoretical curve was obtained as a sum over the final vibrational resonant cross sections, in the resonance region ($\sim2$~eV), and the non-resonant background contribution $\sigma^{bg}(\epsilon)$, as: \be \sigma^{tot}(\epsilon) = \sum_{v'} \sigma_{0\rightarrow v'}^{res}(\epsilon) + \sigma^{bg}(\epsilon)\,,\ee where $\sigma^{bg}(\epsilon)$ was calculated using the \textit{R}-matrix method to evaluating the \textit{T}-matrix at the equilibrium internuclear distance for all symmetries except the resonant $^2\Pi_g$ one. Above $\sim2.5$~eV our cross sections are in good agreement with the observed ones. At lower energies the background is too high. This is due to the lack of polarization effects in the \textit{R}-matrix calculation which become increasingly important at very low collision energies. It is possible to include these effects in the calculation but only by making the calculation significantly more expensive~\cite{jtCH4}. In the resonance region below $\sim2.5$~eV, the theoretical peaks are higher than the experimental ones and show a  small shift which is, however, not present in the comparison with the inelastic cases shown in Figs.~\ref{fig:Inelastic_comp_Allan} and \ref{fig:comp_vicic}.

There is an important difference between the resonance widths shown in Fig.~\ref{fig:pes}(b). The width calculated \textit{ab initio} using the \textit{R}-matrix method (solid line), shows a monotonic increase toward short bond-lengths which contrasts with the bell-shaped behavior of the semi-empirical $\G(R)$ obtained in I (dashed line). However, the two curves overlap for bond-lengths greater than 1.8 a$_0$, which covers the Franck-Condon region for the transitions from the vibrational ground state. This explains the absence of any substantial differences  between the present calculations and those reported in I for  the corresponding vibrational excitation cross sections.

\begin{center}
\begin{figure}
\includegraphics[scale=.7]{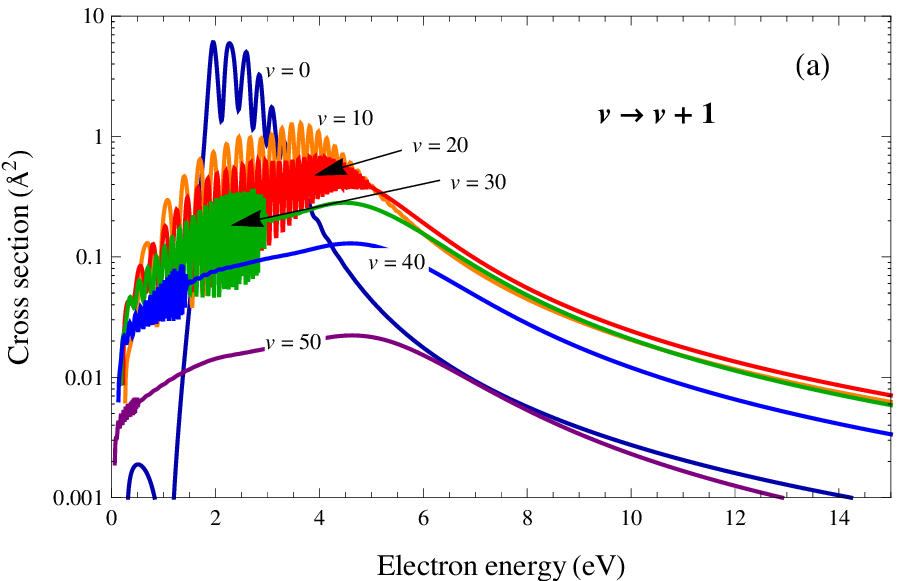}\hspace{1cm}
\includegraphics[scale=.7]{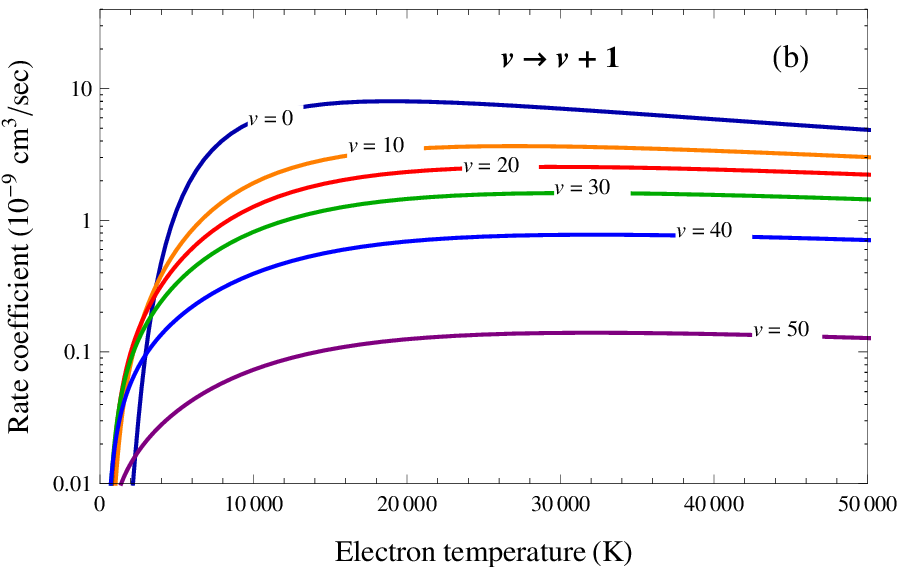}
\caption{(a) electron-N$_2$ resonant vibrational-excitation cross sections and
(b) corresponding rate coefficients for selected single-quantum transitions. \label{fig:VExsec}}
\end{figure}
\end{center}

Figure~\ref{fig:VExsec} shows a set of the new calculated RVE cross
sections and the corresponding rate coefficients for the $v\rightarrow
v+1$ single-quantum transitions, which play a prominent role in the
plasma vibrational kinetic. The cross section curves displayed  are
labeled with the selected values of the vibrational quantum number
$v$. As extensively discussed in paper I the sharp
  peaks in the cross sections shown in
  Figs.~\ref{fig:Inelastic_comp_Allan}--\ref{fig:VExsec} are due to
  the well know boomerang-oscillations  and their position corresponds
  to the energy of the resonant vibrational levels. For high energies
  the cross sections drop down of several orders of magnitude, relative to
  their peak value, so that in the integration of the rate
  coefficients in Eq.~(\ref{eq:rate coefficient}) they were considered
  negligible beyond 15~eV.

\begin{center}
\begin{figure}
\includegraphics[scale=.2, angle=-90]{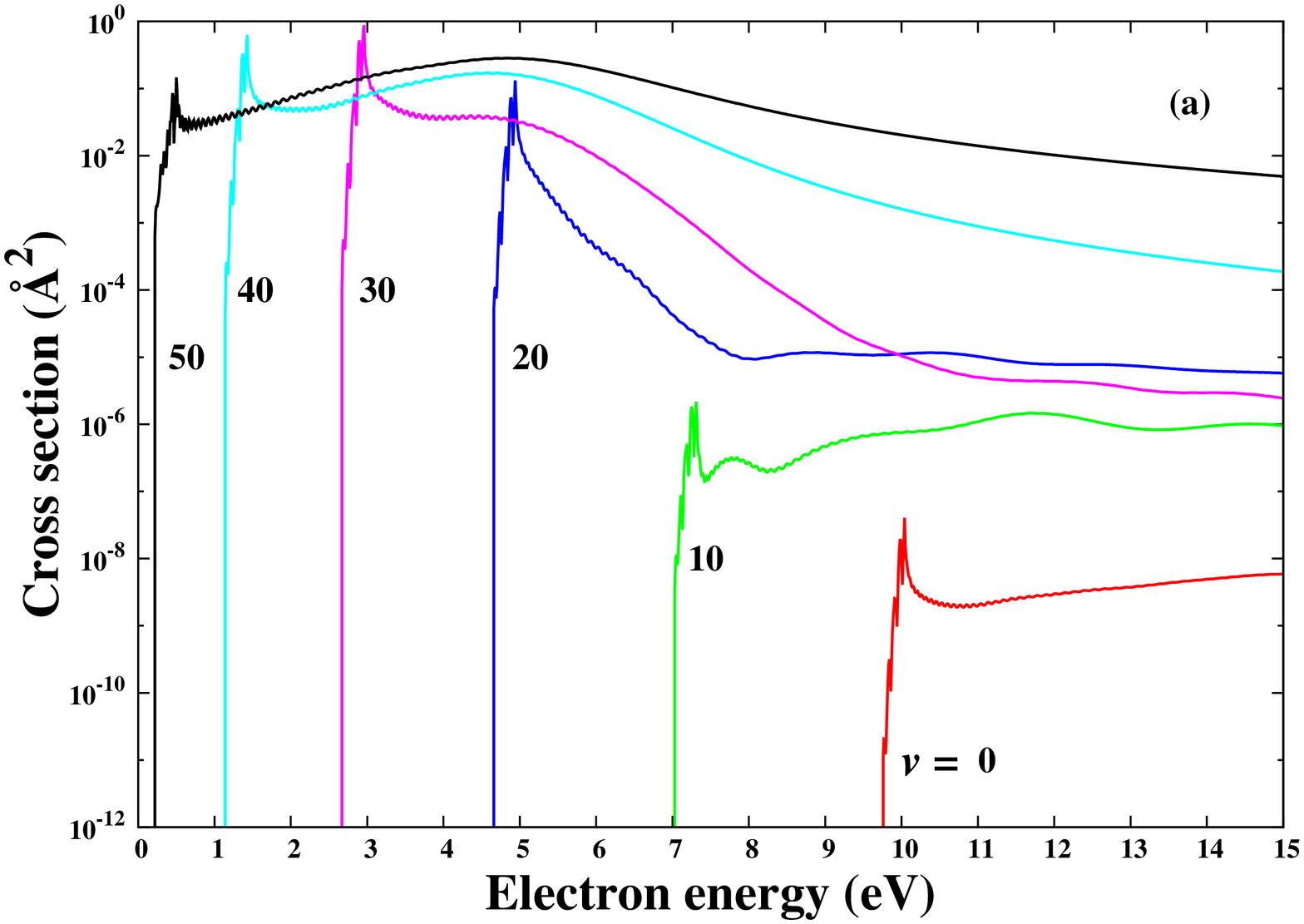}
\includegraphics[scale=.2, angle=-90]{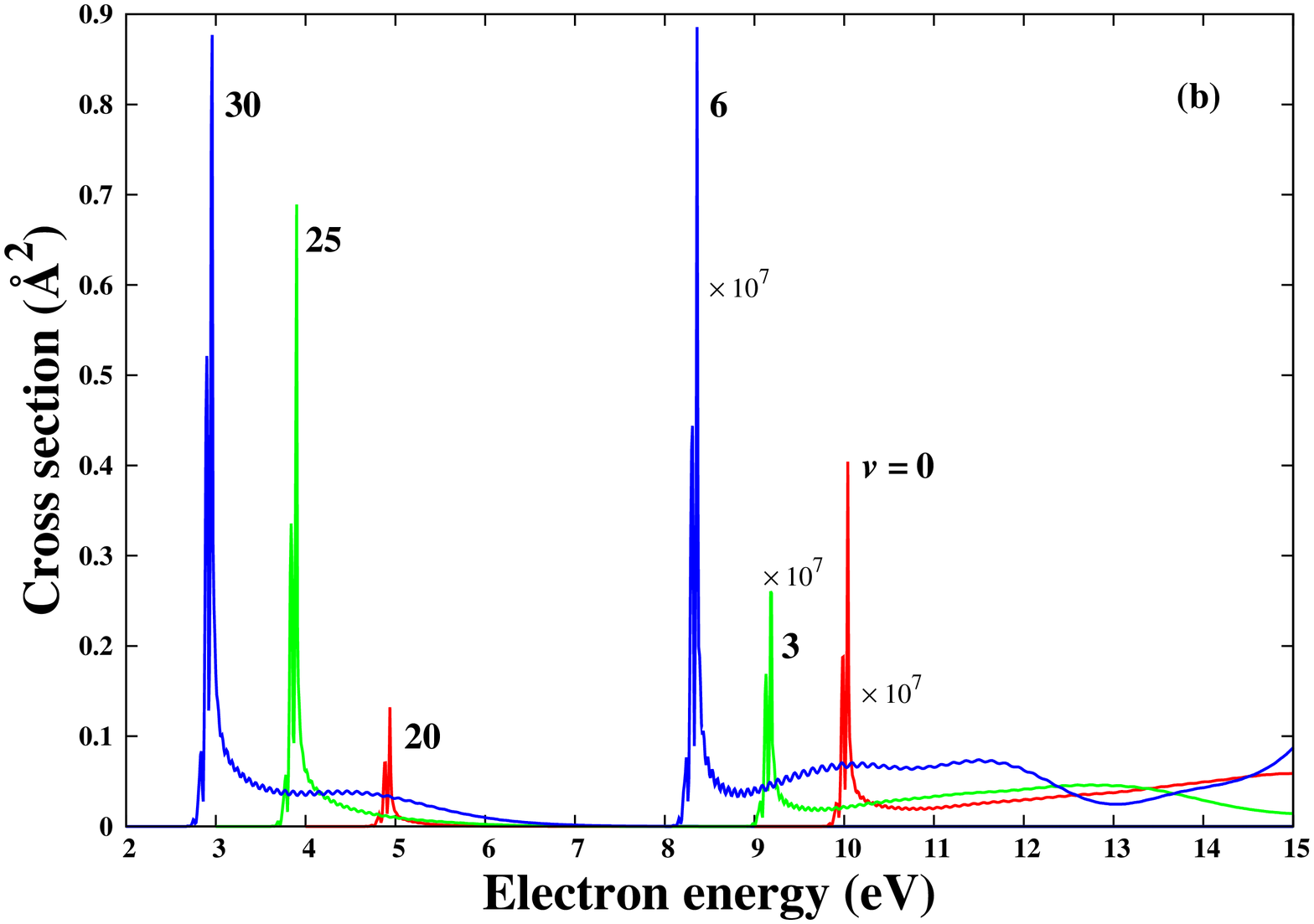}
\includegraphics[scale=.2, angle=-90]{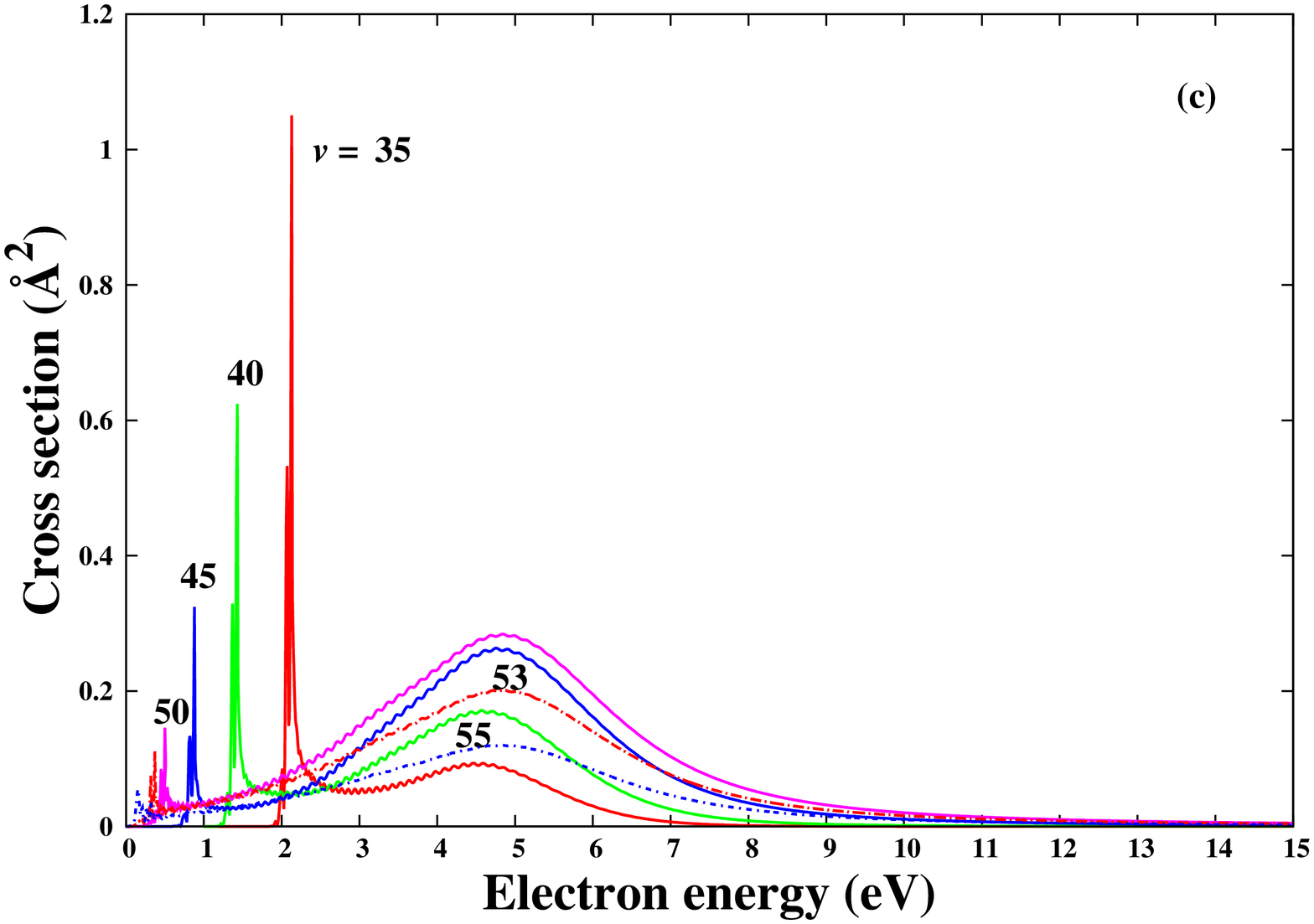}
\caption{Resonant electron-N$_2$ dissociation cross sections for some initial vibrational levels, as indicated in the figure, and for   $J=0$. Panel (a) gives plots on a log \textit{y}-scale. In panel (b) the cross sections for $v=0,3$ and 6 have been multiplied by   $10^7$, as shown. In panel (c) the dashed lines ($v$ = 53 and 55)   indicate a decreasing trend for the corresponding cross sections.  \label{fig:DISSxsec}}
\end{figure}
\end{center}

The cross sections for the dissociative process
(\ref{eq:Rdiss_process}), calculated by Eq.~(\ref{eq:cont_xsec}), are
shown in Fig.~\ref{fig:DISSxsec} for some initial vibrational levels
as a function of the incident electron energy. An immediate result
which can be drawn from Fig.~\ref{fig:DISSxsec}a, where the plots are
represented in log \textit{y}-scale, is that the cross sections for
low levels ($v \lesssim 20$) are extremely small, so that the role of
the corresponding dissociative processes in plasma kinetics can be
expected to be negligible. For higher levels however, the cross
sections tend to saturate above $10^{-2}$ \AA$^2$, which implies that
in strong non-equilibrium plasma conditions, that is when the higher
levels are overpopulated with respect to the Boltzmann distribution,
the dissociative processes, starting from these levels, will play a
major role. Figure~\ref{fig:DISSxsec}b shows the cross sections on a
linear scale for some $v$ values ranging from 0 through 30. All the
curves exhibit two large, sharp peaks and a structure of smaller
intensity very close to the apparent threshold. These features are
also present in all the calculated cross sections as can be seen in
Fig.~\ref{fig:DISSxsec}c. Inspection of the numerical values shows
that the peak positions coincide with some of the vibrational
eigenvalues of the N$_2^-(\textrm{X}\ ^2\Pi_g)$ resonant state, placed
inside the electronic affinity gap, of $|0.30|$ eV (see Table~\ref{tab:spectr_data}), between the
asymptotic limits of the N$_2(\textrm{X}\ ^1\Sigma_g)$ and
N$_2^-(\textrm{X}\ ^2\Pi_g)$ potential curves. In this interval we
found twelve resonant vibrational levels which are the only ``bound''
states that can lead to dissociation. These levels, in fact, can enter
in resonance with the continuum of the N$_2$ ground state while above
the N$_2^-$ dissociation limit, instead, the interaction of the two
continua occurs.

As $v$ is increased the threshold of the resonant dissociation process
is lowered so that the cross section peaks in Fig.~\ref{fig:DISSxsec}b
move toward lower energies. On the other side of the peaks and for
large incident energies, some oscillation is observed in the curves.
This is due to numerical noise. In fact, the very small absolute
values of the low-level cross sections ($\sim 10^{-8}$ \AA$^2$)
implies a reduced accuracy in the calculations so that numerical
instabilities become evident. For $v=8,9,10$ the corresponding cross
sections ($\sim 10^{-7}$ \AA$^2$, not shown) are even larger while for
$v\gtrsim20$, for which the cross sections become significantly large,
the oscillations remain confined to very small values and disappear
from the plots.

Figure~\ref{fig:DISSxsec}c shows cross sections for some initial
levels with $v\geq 35$; besides the sharp peaks already discussed,
there is also a broad maximum arising about 5~eV,
whose intensity grows with the vibrational quantum number up to $v=50$
then decreases for $v=53$ and 55 (dashed lines). This maximum occurs when the N$_2$ vibrational states
lie in the continuum spectrum of N$_2^-$ ion and is therefore due
to the interplay of the Franck-Condon overlap between the N$_2$ and N$_2^-$ bound
and continuum levels respectively, during the electron capture, and
that of the two continua in the emission process.

\begin{center}
\begin{figure}
\includegraphics[scale=.25, angle=-90]{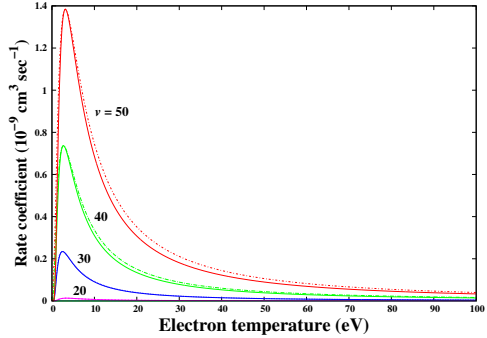}
\caption{Calculated (full lines) and fitted (dashed lines) rate  coefficients for resonant dissociation as a function of the electron temperature and for the indicated vibrational levels. \label{fig:ratefit}}
\end{figure}
\end{center}

Rate coefficients were also calculated for process
(\ref{eq:Rdiss_process}) using Eq.~(\ref{eq:rate coefficient}) and the
corresponding dissociative cross sections. Figure~\ref{fig:ratefit}
shows the rates as a function of the electron temperature for the
initial vibrational levels $v = 20, 30 ,40$ and 50. These
rates follow the same trend as the cross sections, being negligible
for low $v\ (\lesssim 20)$ and becoming  significant  for
higher vibrational levels. Likely in the RVE case, the
  dissociation cross sections decrease rapidly as a function of the
  electron energy, so again we extended the
  integration in the calculation of the rate coefficients up to
  15~eV.

The rates can be easily and accurately reproduced using the following analytical expression:
\begin{equation}
\k_{v}(T)=\k_{v}^{max}\left(\frac{T_{v}^{max}}{T}\right)^{3/2}e^{-\frac{T_{v}^{max}}{T}}\,, \label{eq:ratefit}
\end{equation}
already successfully tested for electron-H$_2$ scattering~\cite{Celiberto_et_al_2013_H2}. The equation requires only two parameters, $T_{v}^{max}$ and $\k_{v}^{max}$, which are the coordinates of the maximum value of the calculated rates for each level $v$. Eq.~(\ref{eq:ratefit}) works quite well for all $v$ but the last three values ($v=56,57$ and 58) as the corresponding rates show some irregularity, probably coming from the related cross sections which, for these very high levels, suffer from reduced numerical accuracy. $T_{v}^{max}$ and $\k_{v}^{max}$ are provided here as a supplementary material for both dissociative ($0\leq v\leq 55$) and RVE processes ($0\leq v \leq v'\leq 55$). The RVE rates for $v\geq v'$ can be obtained by detailed balance~\cite{Celiberto_et_al_PPCF_2012_BeH(+)}.

\begin{center}
\begin{figure}
\includegraphics[scale=.55]{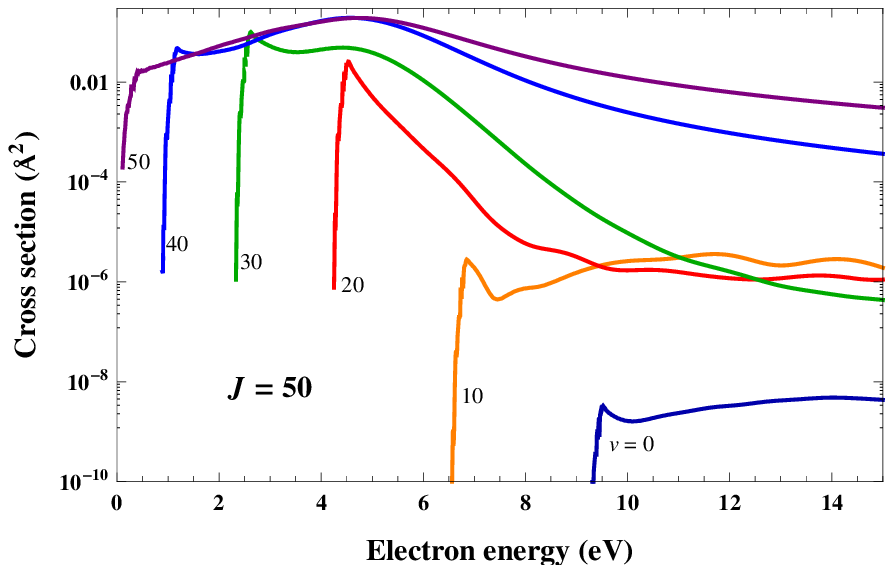}
\includegraphics[scale=.55]{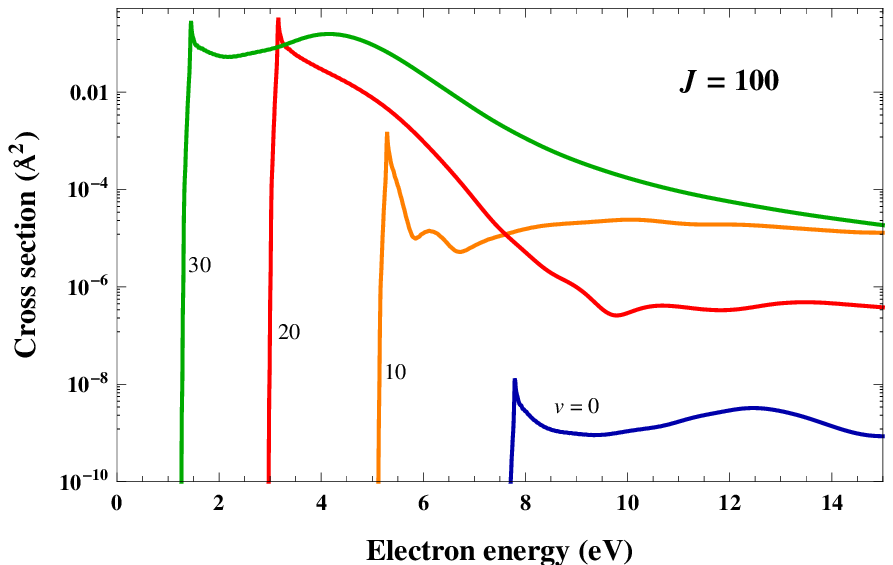}
\includegraphics[scale=.55]{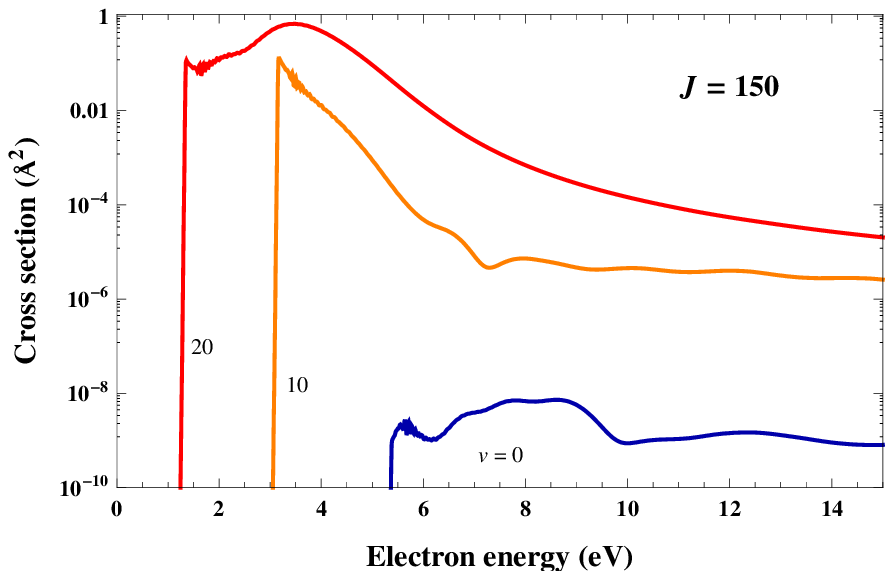}
\caption{Resonant dissociation cross sections as a function of the electron energy, for some vibrational levels and for $J=50$, $J=100$ and $J=150$. \label{fig:xsec-J}}
\end{figure}
\end{center}

We also investigated the behavior of the dissociative process as a
function of rotational state. Figure~\ref{fig:xsec-J} shows the cross
sections for $J=50, 100$ and 150 and for different $v$. In calculating
these cross sections we have started the integration in
Eq.~(\ref{eq:cont_xsec}) above the centrifugal barrier created in the
N$_2$ potential curve by the nuclear rotation, instead of from the
dissociation energy $\e_{th}$. This barrier, in fact, can support a
number of quasi-bound states which can lead to dissociation by
tunneling. We have assumed that the contribution of these metastable
states, if they exist, to dissociation is small compared to the
process occurring from the repulsive part of the potential curve, due
to the delay accumulated by the nuclei inside the barrier.
Figure~\ref{fig:xsec-J} shows, for a given $J$, that again the cross
sections present structures near the threshold and become of
significant values for high vibrational levels, comparable in
magnitude with those for $J=0$.

\section{Conclusions \label{sec:conc}}
In this article we update the resonant vibrational excitation cross sections and rate coefficients reported in I. Our new calculations use accurate potential energies for both N$_2$ and N$_2^-$ ground states. For the neutral molecule we use the experimentally-derived potential curves~\cite{:/content/aip/journal/jcp/125/16/10.1063/1.2354502}, while for N$_2^-$ resonant state we perform new calculations using the \textit{R}-matrix method, obtaining also the resonance width as a function of the bond length.

We extend the cross section calculations to the study of dissociative
resonant vibrational excitations using the same model. The
energy-dependent cross section curves obtained show some sharp peaks
close to the process threshold, which are caused by resonant bound
states located in between the dissociation limits of the N$_2$ and
N$_2^-$ molecules. Above these levels resonant coupling occurs among
the two continua of the neutral and ionic species. The dissociation
cross sections are of very small values for low initial vibrational
levels, but become important for high $v$ ($\gtrsim 20$).  The same
behavior is shown by the corresponding rate coefficients. For these
last quantities, as well as for the RVE rates, a two-parameter
analytical fitting expression has been formulated for their rapid and
accurate evaluation which should be useful for practical applications.
Finally, cross sections for $J=50, 100$ and 150, and for different $v$
have been also investigated. Their behavior, as well as their order of
magnitude, is comparable to the cross sections for $J=0$.

As already stressed previously, the resonant dissociative cross sections starting from $v=0$ level is particularly small. This can be better seen in comparison with the experimental measurements of $\textrm{N}_2(v=0)$ reported by Cosby~\cite{:/content/aip/journal/jcp/98/12/10.1063/1.464385} which found that the dominant contribution to dissociation comes from electronically excited states leading to N($^2D$) + N($^4S$). However, in nitrogen plasmas, exposed to high temperature and  electric fields, the contribution coming from the resonant dissociation from all the vibrational levels, is comparable with the dominant kinetic mechanism of dissociation induced by vibrational quanta exchange, which gradually excite the molecules up to the vibrational continuum (the so-called \emph{`pure-vibrational mechanism'}) as discussed in Ref.~\cite{capitelliN2PVM}.\\

The complete set of cross sections and the corresponding rate constants can be downloaded from the `Phys4Entry' database~\cite{F4Edatabase} and the parameters of the fit in Eq.~(\ref{eq:ratefit}) can be found as supplementary material of this paper.

\section*{Acknowledgements}
This work received funding from the European Community's Seventh Framework Programme (FP7/2007-2013) under grant agreement n.~242311. One of the authors (DAL) would like to also thank support from Themisys Limited for supporting a studentship. The authors wish to thank Drs. W.M.~Huo, R.~Jaffe and D.W.~Schwenke (NASA Ames Research Center) for providing the N$_2$ potential energy curve in Ref.~\cite{:/content/aip/journal/jcp/125/16/10.1063/1.2354502}.

\section*{References}
\addcontentsline{toc}{section}{References}

\bibliographystyle{is-unsrt}{}  


\end{document}